\documentclass[cameraready]{Interspeech}

\title{BASENet: Band-Adapted Speech Enhancement Network with Cross-Band Attention}

\author[affiliation={1}, correspondingauthor]{Damien}{MARTINS GOMES}
\author[affiliation={1}]{Francois}{CAPMAN}

\address{
    $^1$ Thales SIX GTS, FRANCE 
}

\email{damien.martins-gomes@thalesgroup.com, francois.capman@thalesgroup.com}


\usepackage{comment}
\usepackage{float} 
\usepackage{graphicx}  
\usepackage[table]{xcolor}  
\usepackage{pifont}

\begin{document}

\maketitle

\begin{abstract}
Speech enhancement models typically apply uniform capacity across all frequencies, disregarding the non-uniform spectral resolution of human hearing.
We propose BASENet, a frequency-adapted architecture that partitions the spectrum into Bark-scale bands and assigns each a scaled-capacity encoder derived from critical-band density, automatically granting deeper branches to perceptually dense low frequencies and lighter ones to high frequencies. A cross-band attention module captures harmonic dependencies across bands through compact frequency-pooled representations at linear complexity. Built on inverted residual blocks with dense connectivity and a convolutional recurrent network, BASENet achieves PESQ~3.55 and STOI~96\% on VoiceBank+DEMAND with only 0.83\,M parameters and 7.3\,G~MACs, the fewest parameters among all methods with PESQ~$\geq$\,3.50. A causal variant (PESQ~3.44) surpasses several non-causal baselines, confirming suitability for real-time streaming on resource-constrained devices.
\end{abstract}

\section{Introduction}

Single-channel speech enhancement aims to recover clean speech from noisy observations, with applications in hearing aids, voice communication, and ASR front-ends~\cite{loizou2013speech}. Deep learning approaches have progressed from masking-based methods~\cite{wang2014training} to architectures jointly estimating magnitude and phase~\cite{yin2020phasen,lu2023mpsenet} or operating on complex spectrograms~\cite{hu2020dccrn,tan2022dptfsnet}. State-of-the-art models such as MH-SENet~\cite{kim25q_interspeech}, Mamba-SEUNet~\cite{wang2025mambaseunetmambaunetmonaural}, and MP-SENet~\cite{lu2023mpsenet} achieve impressive quality but at substantial computational cost, while generative approaches based on diffusion~\cite{richter2023sgmse} or flow matching~\cite{liu2024speechflow} require iterative sampling that limits real-time deployment. Moreover, many high-performing architectures rely on bidirectional or non-causal operations that preclude streaming inference, a critical requirement for hearing aids and live communication. A key limitation shared by many architectures is their uniform treatment of the frequency axis. Speech exhibits highly non-uniform spectral characteristics governed by the auditory system's frequency resolution: low-frequency critical bands are narrow yet perceptually dense, encoding fundamental frequency and harmonic structure; mid-frequency bands carry formants critical for intelligibility; and high-frequency bands span wider ranges with lower perceptual acuity~\cite{moore2012introduction}. Standard full-band encoders ignore these differences, applying equal capacity across the spectrum. Recent sub-band approaches have begun to address this. FullSubNet~\cite{hao2021fullsubnet} and extensions~\cite{chen2023intersubnet} fuse full-band context with sub-band processing but apply uniform capacity per sub-band. DeepFilterNet~\cite{schroter2022deepfilternet} exploits psychoacoustic structure for the \emph{signal representation} but not the \emph{model architecture}. Band-Split RNN~\cite{yu23b_interspeech} introduced explicit sub-band splitting with interleaved band- and sequence-level modelling, yet uses uniform depth across sub-bands and relies on shared BLSTMs for cross-band modelling without leveraging perceptual importance to modulate encoder capacity. A common tension persists: existing methods either employ uniform-capacity sub-band networks~\cite{hao2021fullsubnet,yu23b_interspeech} or resort to expensive full-band self-attention~\cite{wang2023tfgridnet}, leaving open how to \emph{systematically} derive capacity from auditory principles while maintaining cross-band coherence.\\
We propose \textbf{BASENet} (Band-Adapted Speech Enhancement Network), a frequency-adapted architecture whose encoder capacity is a closed-form function of auditory bandwidth. BASENet partitions the spectrum into $B$ perceptually motivated bands along the Bark scale~\cite{zwicker1961subdivision} and computes a per-band \emph{critical-band density} measuring how many Bark bands are packed per Hertz. This density directly determines encoder depth: narrow, perceptually dense low-frequency bands automatically receive deeper, wider branches, while broad high-frequency bands are processed by lighter ones. To restore cross-band coherence, a compact cross-band attention mechanism operates on frequency-pooled band summaries, enabling efficient inter-band information exchange at $\mathcal{O}(NF_bB)$ complexity. The architecture combines MobileNetV3-inspired inverted residual blocks~\cite{howard2019searching} with dense connectivity~\cite{huang2017densely} and a Convolutional Recurrent Network (CRN)~\cite{tan2019learning} for temporal modelling. Crucially, all components---including the cross-band attention---operate frame-by-frame, enabling native causal streaming by simply replacing the bidirectional GRU with a unidirectional variant. Our contributions are:
\begin{itemize}
    \item A \textbf{perceptually scaled encoder} that derives per-band capacity from Bark-scale critical-band density via a single closed-form rule, eliminating per-band hyperparameter tuning.
    \item A \textbf{cross-band attention} module that models harmonic and formant dependencies across bands through compact frequency-pooled representations at linear complexity.
    \item A \textbf{lightweight, streaming-ready architecture} (0.83\,M parameters, 7.3\,G MACs) that achieves the highest PESQ among all sub-1\,M-parameter methods with PESQ~$\geq$\,3.50, while natively supporting causal inference for real-time deployment.
\end{itemize}

\begin{figure*}[t]
    \centering
    \includegraphics[width=\textwidth]{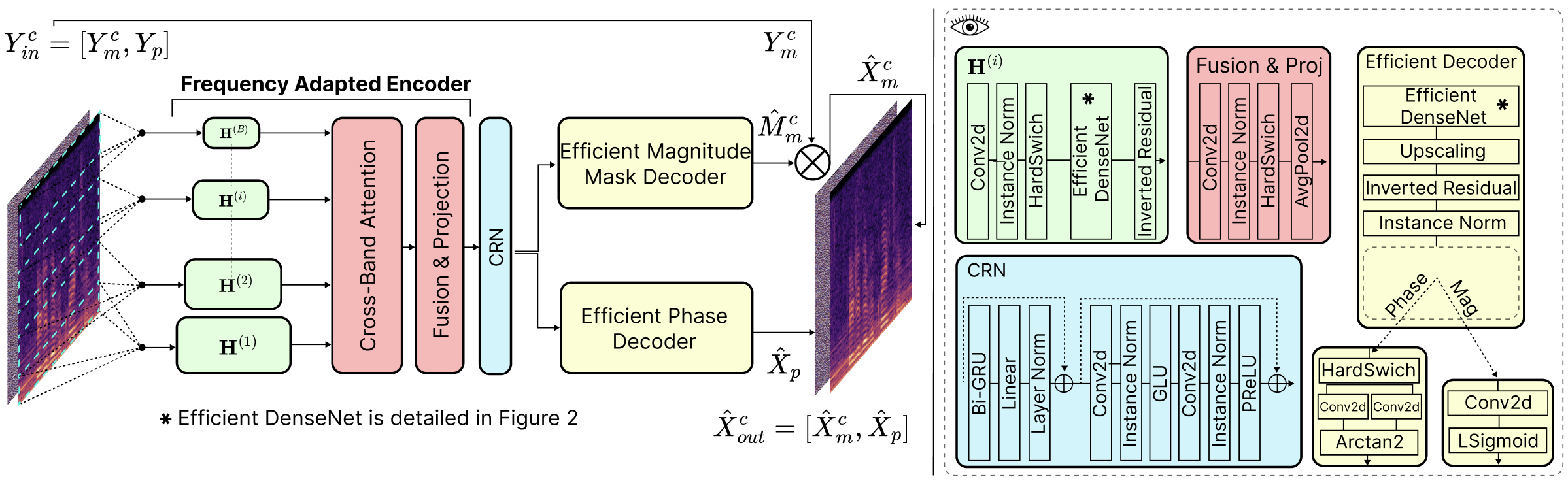}
    \caption{Overview of BASENet. Noisy magnitude and phase spectrograms are concatenated and processed by a frequency-adapted encoder that partitions the spectrum into $B$ Bark-scale bands with capacity scaled by perceptual density. Cross-band attention enables inter-band information exchange before CRN temporal modelling. Decoders predict the magnitude mask and enhanced phase.}
    \label{fig:architecture}
\end{figure*}
\section{Proposed Method}
\label{sec:method}

\subsection{Overall Architecture}
\label{sec:overall}

An overview of BASENet is shown in Fig.~\ref{fig:architecture}. We adopt the mask-and-phase estimation paradigm of MP-SENet~\cite{lu2023mpsenet}: given a noisy speech signal $\mathbf{y} \in \mathbb{R}^{T}$, we compute the Short-Time Fourier Transform (STFT) $\mathbf{Y} \in \mathbb{C}^{F\times N}$ to obtain magnitude $\mathbf{Y}_m=|\mathbf{Y}|$ and phase $\mathbf{Y}_p=\angle\mathbf{Y}$ spectrograms, where $F=\lfloor n_{\text{fft}}/2\rfloor+1$ frequency bins and $N$ time frames. A power-law compression $\mathbf{Y}_m^c=|\mathbf{Y}|^{c}$ with $c = 0.3$ reduces dynamic range. Compressed magnitude and phase are concatenated along the channel dimension to form the input $\mathbf{X} \in \mathbb{R}^{2\times N\times F}$.\\
The input is processed by a \emph{Frequency-Adapted Encoder} (Sec.~\ref{sec:encoder}), which partitions the spectrum into $B$ perceptually motivated bands and applies scaled-capacity processing proportional to perceptual importance. A \emph{Cross-Band Attention} module (Sec.~\ref{sec:crossband}) then enables inter-band information exchange. The fused representation is passed to a CRN for temporal modelling. Throughout the architecture, we employ efficient inverted residual blocks with dense connectivity (Sec.~\ref{sec:efficient}) to minimise computational cost while preserving representational capacity. Finally, lightweight magnitude and phase decoders predict a time-frequency mask $\hat{\mathbf{M}}_m$ and enhanced phase $\hat{\mathbf{X}}_p$. The enhanced complex spectrogram is reconstructed as:
\begin{equation}
    \hat{\mathbf{S}} = \underbrace{\mathbf{Y}_m\odot\hat{\mathbf{M}}_m}_{\hat{\mathbf{X}}_m}\odot e^{j\hat{\mathbf{X}}_p},
\end{equation}
where $\odot$ denotes element-wise multiplication. We follow the same training procedure and loss functions as MP-SENet~\cite{lu2023mpsenet}, and refer the reader to that work for details.

\subsection{Frequency-Adapted Encoder}
\label{sec:encoder}
\textbf{Perceptually Motivated Frequency Decomposition.}
The human auditory system resolves spectral detail non-uniformly: low-frequency regions are analysed through narrow critical bands that provide fine resolution, while high-frequency regions are covered by progressively wider bands with coarser resolution~\cite{moore2012introduction}. The Bark scale~\cite{zwicker1961subdivision} formalises this observation by mapping each frequency $f$ (in~Hz) to a position $z(f)$ on a perceptual scale (in~Bark) via\footnote{Approximation due to Traunm\"uller~\cite{traunmuller1990analytical}. One Bark corresponds to one critical bandwidth; the total audible range spans roughly 0--24~Bark.}:
\begin{equation}
    z(f)=13\arctan(0.00076\,f)+3.5\arctan\left(\frac{f}{7500}\right)^{2}.
    \label{eq:bark}
\end{equation}
Because critical bands are narrow at low frequencies and wide at high frequencies, the number of Bark units per Hertz---i.e., the \emph{critical-band density}---decreases with frequency. We embed this perceptual non-uniformity directly into the encoder architecture. The input spectrogram is partitioned into $B$ non-overlapping frequency bands $\mathcal{B}=\{[f_0, f_1),\,[f_1, f_2),\,\ldots,\,[f_{B-1}, f_B]\}$ with $f_0 = 0$ and $f_B = f_s/2$, whose boundaries are placed at transitions between perceptually distinct spectral regions (fundamental/harmonic, formant, fricative), guided by the Bark scale and validated empirically in Sec.~\ref{sec:ablation}. For each band~$b$, the critical-band density is:
\begin{equation}
    \rho_b = \frac{z(f_{b}) - z(f_{b-1})}{f_{b} - f_{b-1}},
    \label{eq:cbd}
\end{equation}
which measures how many Bark units are spanned per Hertz within band~$b$. A higher $\rho_b$ indicates finer auditory resolution and, consequently, greater sensitivity to spectral distortion.\\
\textbf{Density-Driven Capacity Allocation.}
The key design principle of BASENet is that encoder capacity should mirror perceptual density: bands where the ear is most sensitive receive deeper processing, while bands with coarser resolution are handled by shallower branches. We map $\rho_b$ to a per-band depth:
\begin{equation}
    L_b = \left\lfloor L_{\max}\cdot
    \frac{\rho_b}{\max_{b'}\rho_{b'}} \right\rceil,
    \label{eq:depth}
\end{equation}
where $L_{\max}$ is the maximum branch depth and $\lfloor\cdot\rceil$ denotes rounding to the nearest integer. The normalisation ensures that the perceptually densest band always receives $L_{\max}$ layers, while remaining bands receive depth proportional to their relative density. This provides a single-parameter ($L_{\max}$) design rule that, given any sampling rate and band partition, produces a scaled-capacity allocation without per-band tuning.\\
\textbf{Scaled-Capacity Band Processing.}
Let $\mathbf{X}^{(b)} \in \mathbb{R}^{C\times N\times F_b}$ denote the $b$-th band after an initial $1\times1$ projection to $C$ channels, where $F_b$ is the number of frequency bins in band~$b$. Each band is processed by a dedicated encoder branch of depth $L_b$:
\begin{equation}
    \mathbf{H}^{(b)} = \mathcal{E}_{b}\left(\mathbf{X}^{(b)}\right),
    \quad \text{depth}(\mathcal{E}_b) = L_b,
    \label{eq:band_proc}
\end{equation}
where $\mathcal{E}_{b}$ consists of a DenseBlock of $L_b$ layers followed by an inverted residual block (Sec.~\ref{sec:efficient}). Deeper branches (high $L_b$) develop larger receptive fields along the frequency axis through exponentially dilated convolutions, enabling fine-grained modelling of harmonics and pitch structure in perceptually critical low-frequency regions. Shallower branches (low $L_b$) provide sufficient capacity for the spectrally smoother high-frequency content while keeping computation low.
\begin{figure}[h]
    \centering
    \includegraphics[width=\columnwidth]{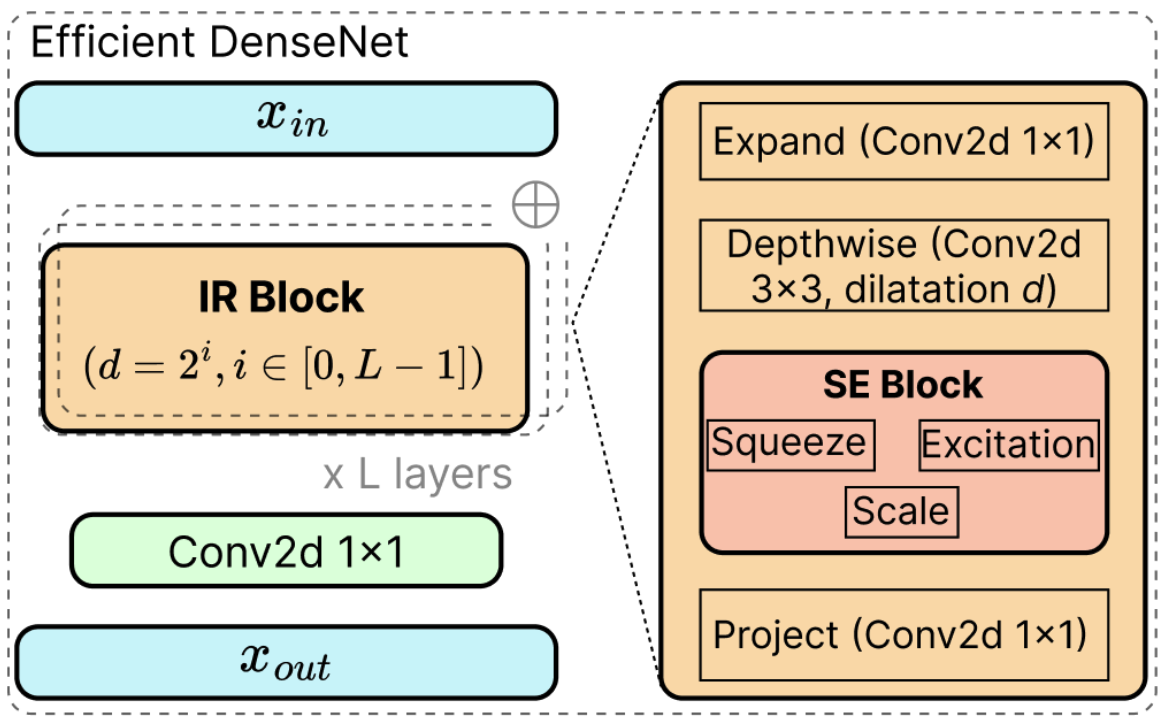}
    \caption{Efficient DenseNet architecture with $L$ inverted residual (IR) blocks using exponential frequency dilation ($d = 2^i$). Each IR block applies expansion, dilated depthwise convolution, squeeze-excitation, and projection.}
    \label{fig:efficient}
\end{figure}
\vspace{-5mm}
\subsection{Cross-Band Attention}
\label{sec:crossband}

Although bands are processed independently, speech production induces strong inter-band dependencies: harmonics generated by the glottal source span the entire spectrum, and formant transitions create correlated energy modulations across bands~\cite{rabiner1978digital}. To model these dependencies efficiently, we introduce \emph{cross-band attention} operating on compact band-level summaries rather than full time-frequency grids. Let $d_k = C/r_a$ denote the attention projection dimension, where $r_a$ is a reduction ratio (we use $r_a = 4$). For each band~$b$, we compute query, key, and value projections:
\begin{align}
    \mathbf{Q}^{(b)} &= \mathbf{W}_Q^{(b)}\,\mathbf{H}^{(b)}
        \in\mathbb{R}^{d_k\times N\times F_b}, \label{eq:Q}\\
    \mathbf{K}^{(b)} &= \mathbf{W}_K^{(b)}\,\mathbf{H}^{(b)}
        \in\mathbb{R}^{d_k\times N\times F_b}, \label{eq:K}\\
    \mathbf{V}^{(b)} &= \mathbf{W}_V^{(b)}\,\mathbf{H}^{(b)}
        \in\mathbb{R}^{C\times N\times F_b}. \label{eq:V}
\end{align}
Rather than allowing each time-frequency bin to attend to all others---incurring prohibitive $\mathcal{O}(N^2 F^2)$ cost---we form a compact global context by averaging keys and values across frequency bins within each band, then concatenating across all $B$ bands:
\begin{align}
    \bar{\mathbf{K}} &=
    \operatorname{Concat}\left[\operatorname{AvgPool}_F\left(\mathbf{K}^{(b)}\right)\right]_{b=1}^{B}
     \in \mathbb{R}^{d_k\times N\times B},
    \label{eq:global_k}\\
    \bar{\mathbf{V}} &=
    \operatorname{Concat}\left[\operatorname{AvgPool}_F\left(\mathbf{V}^{(b)}\right)\right]_{b=1}^{B}
     \in \mathbb{R}^{C\times N\times B}.
    \label{eq:global_v}
\end{align}
Each band then attends to this shared context. Operating independently per time frame~$n$, we compute:
\begin{equation}
    \mathbf{A}^{(b)}_n =
    \operatorname{softmax}\left(\frac{{\mathbf{Q}^{(b)}_n}^{\top}\bar{\mathbf{K}}_n}
    {\sqrt{d_k}}\right)\bar{\mathbf{V}}_n^{\top},
    \label{eq:attn}
\end{equation}
where ${\mathbf{Q}^{(b)}_n}^{\top} \in \mathbb{R}^{F_b\times d_k}$ multiplied by $\bar{\mathbf{K}}_n \in \mathbb{R}^{d_k\times B}$ yields attention weights in $\mathbb{R}^{F_b\times B}$, applied to $\bar{\mathbf{V}}_n^{\top} \in \mathbb{R}^{B\times C}$ to produce the output in $\mathbb{R}^{F_b\times C}$. The final output is:
\begin{equation}
    \tilde{\mathbf{H}}^{(b)} = \mathbf{H}^{(b)}+
    \mathbf{W}_O^{(b)}\,\mathbf{A}^{(b)}.
\end{equation}
This achieves $\mathcal{O}(NF_bB)$ complexity per band, making it tractable even for large $F$ and fully compatible with real-time causal inference since each frame is processed independently.
\subsection{Efficient DenseNet Architecture}
\label{sec:efficient}
Each encoder and decoder branch employs a lightweight architecture combining inverted residual blocks~\cite{howard2019searching} with dense connectivity~\cite{huang2017densely}, as illustrated in Fig.~\ref{fig:efficient}. The core inverted residual (IR) block is:
\begin{equation}
    \mathrm{IR}(\mathbf{x}) = \mathbf{x} +
        \mathrm{Proj}\bigl(\mathrm{SE}(\mathrm{DWConv}(\mathrm{Expand}(\mathbf{x})))\bigr),
\end{equation}
where Expand projects $C$ channels to $rC$ ($r \in \{2,4\}$), DWConv applies depthwise separable convolution with kernel $3 \times 3$ and dilation $d$, SE denotes squeeze-and-excitation~\cite{hu2018squeeze}, and Proj reduces back to $C$ channels. This achieves $\mathcal{O}(K^2 rC+rC^2)$ complexity versus $\mathcal{O}(K^2 C^2)$ for standard convolutions. We use Hardswish activation for efficient inference. Dense connectivity promotes feature reuse: $\mathbf{z}_i=\mathrm{IR}_i([\mathbf{z}_0,\ldots,\mathbf{z}_{i-1}])$, with $L$ blocks stacked using exponentially increasing dilation rates $d = 2^i,\;i \in [0,L{-}1]$ along the frequency axis, capturing multi-scale patterns from narrow-band harmonics to broad formants. A bottleneck $1 \times 1$ convolution reduces concatenated features back to $C$ channels.
\subsection{Temporal Modelling and Decoders}
\label{sec:decoders}

After cross-band fusion, band features are concatenated along frequency and processed by a CRN for temporal modelling. Unlike Conformer-based approaches~\cite{lu2023mpsenet,kim2021seconformer} that incur $\mathcal{O}(N^2)$ complexity from self-attention, CRN achieves $\mathcal{O}(N)$ complexity using a uni/bidirectional GRU followed by convolutional layers with GLU activations~\cite{tan2019learning}, enabling efficient frame-by-frame processing. The magnitude and phase decoders follow the same design as MP-SENet~\cite{lu2023mpsenet}: a bounded sigmoid mask for magnitude and an arctan-based projection for phase; we refer the reader to~\cite{lu2023mpsenet} for details.

\begin{table}[h]
\centering
\caption{Comparison on VoiceBank+DEMAND. `--' = unreported. Colours: \textcolor{green}{waveform}, \textcolor{orange}{complex spectrogram}, \textcolor{blue}{magnitude--phase}, \textcolor{red}{magnitude--phase--time} input. \textbf{Bold} = best per column. BASENet achieves the fewest parameters among all methods with PESQ\,$\geq$\,3.50.}
\label{tab:comparison}
\resizebox{\columnwidth}{!}{%
\begin{tabular}{lcccccccc}
\toprule
\textbf{Method} & \textbf{Causal} & \textbf{\#Param.} & \textbf{MACs} & \textbf{PESQ} & \textbf{CSIG} & \textbf{CBAK} & \textbf{COVL} & \textbf{STOI}\,\% \\
\midrule
Noisy & -- & -- & -- &  1.97 & 3.35 & 2.44 & 2.63 & 91 \\
\midrule
\rowcolor{green!15}
DEMUCS~\cite{defossez2020real}  & \ding{55} & 33.5M & -- & 3.07 & 4.31 & 3.40 & 3.63 & 95 \\
\rowcolor{green!15}
SE-Conformer~\cite{kim2021seconformer}  & \ding{55} & -- & -- & 3.13 & 4.45 & 3.55 & 3.82 & 95 \\
\rowcolor{green!15}
MANNER-S~\cite{Shin_2022}  & \ding{55} & 0.90M & \textbf{2.9G} & 3.06 & 4.42 & 3.58 & 3.77 & 95 \\
\midrule
\rowcolor{orange!15}
BSRNN~\cite{yu23b_interspeech} & \ding{55} & 3M & 5.1G & 3.10 & -- & -- & -- & 95 \\
\rowcolor{orange!15}
DPT-FSNet~\cite{tan2022dptfsnet} & \ding{55} & 0.88M & 8.1G & 3.33 & 4.58 & 3.72 & 4.00 & 96 \\
\rowcolor{orange!15}
CMGAN~\cite{cao2022cmgan} & \ding{55} & 1.83M & 20.9G & 3.41 & 4.63 & 3.94 & 4.12 & 96 \\
\midrule
\rowcolor{blue!15}
PHASEN~\cite{yin2020phasen} & \ding{55} & 7.78M & 11.4G & 2.99 & 4.21 & 3.55 & 3.62 & -- \\
\rowcolor{blue!15}
MP-SENet~\cite{lu2023mpsenet} & \ding{55} & 2.05M & 37.2G  & 3.50 & 4.73 & 3.95 & 4.22 & 96 \\
\rowcolor{blue!15}
SE-Mamba~\cite{chao2025investigationincorporatingmambaspeech} & \ding{55} & 2.25M & 32.7G  & 3.55 & 4.77 & 3.95 & 4.26 & 96 \\
\rowcolor{blue!15}
MUSE~\cite{lin24h_interspeech} & \ding{55} & \textbf{0.51M} & 5.2G  & 3.37 & 4.63 & 3.80 & 4.10 & 95 \\
\rowcolor{blue!15}
Mamba-SEUNet~\cite{wang2025mambaseunetmambaunetmonaural} & \ding{55} & 3.78M & 5.2G  & 3.57 & \textbf{4.79} & 4.00 & 4.30 & 96 \\
\rowcolor{red!15}
MH-SENet~\cite{kim25q_interspeech} & \ding{55} & 0.99M & 8.4G  & \textbf{3.62} & \textbf{4.79} & \textbf{4.01} & \textbf{4.34} & \textbf{96} \\
\midrule
\midrule
\rowcolor{blue!15}
\textbf{BASENet-3} & \ding{55} & \underline{0.83M} & 7.3G & 3.55 & 4.65 & 3.95 & 4.18 & \textbf{96} \\
\rowcolor{blue!15}
\textbf{BASENet-3 (Causal)} & \ding{51} & \underline{0.81M} & 7.1G & 3.44 & 4.58 & 3.85 & 4.04 & \textbf{96} \\
\bottomrule
\end{tabular}
}
\end{table}

\vspace{-3mm}
\section{Experiments}

\subsection{Experimental Setup}

\textbf{Dataset.} We evaluate on VoiceBank+DEMAND~\cite{valentini2016investigating}, including 11{,}572 training utterances (28 speakers, SNRs of 0, 5, 10, 15\,dB) and 824 test utterances (2 unseen speakers, SNRs of 2.5, 7.5, 12.5, 17.5\,dB) with mismatched noise conditions.\\
\textbf{Implementation.} We use $n_{\text{fft}}=400$, hop length 100, and 16\,kHz sample rate. Models are trained for 100 epochs on a single NVIDIA Quadro RTX 6000 GPU with batch size~8, Adam optimizer~\cite{kingma2017adam} ($\text{lr}=10^{-3}$, $\beta_1 = 0.9$, $\beta_2 = 0.999$) and exponential decay rate~$0.98$.\\
\textbf{Metrics.} We report wideband PESQ~\cite{rix2001pesq}, STOI~\cite{taal2011stoi}, and composite measures CSIG, CBAK, COVL~\cite{hu2007evaluation}, along with parameter count and MACs.\\
\textbf{Band Configuration.}
We evaluate $B \in \{3,8,12\}$ bands (Table~\ref{tab:ablation}); $B=3$ performs best (PESQ~3.55), with finer partitions degrading quality by fragmenting each branch's spectral context. For $f_s = 16$\,kHz, the $B=3$ boundaries $\mathcal{B} = \{[0, 1),\,[1, 4),\,[4, 8]\}$\,kHz separate fundamental/harmonic (Bark~$0$--$8$, $\rho_{\text{low}} \approx 8.0 \times 10^{-3}$\,Bark/Hz), formant (Bark~$8$--$17$, $\rho_{\text{mid}} \approx 3.0 \times 10^{-3}$), and fricative regions (Bark~$17$--$22$, $\rho_{\text{high}} \approx 1.25 \times 10^{-3}$). Applying Eq.~\eqref{eq:depth} with $L_{\max}=4$ yields depths $L_{\text{low}}=4$, $L_{\text{mid}}=3$, $L_{\text{high}}=2$, directly reflecting the decreasing perceptual density.
\begin{figure}[h]
    \centering
    \includegraphics[width=\columnwidth]{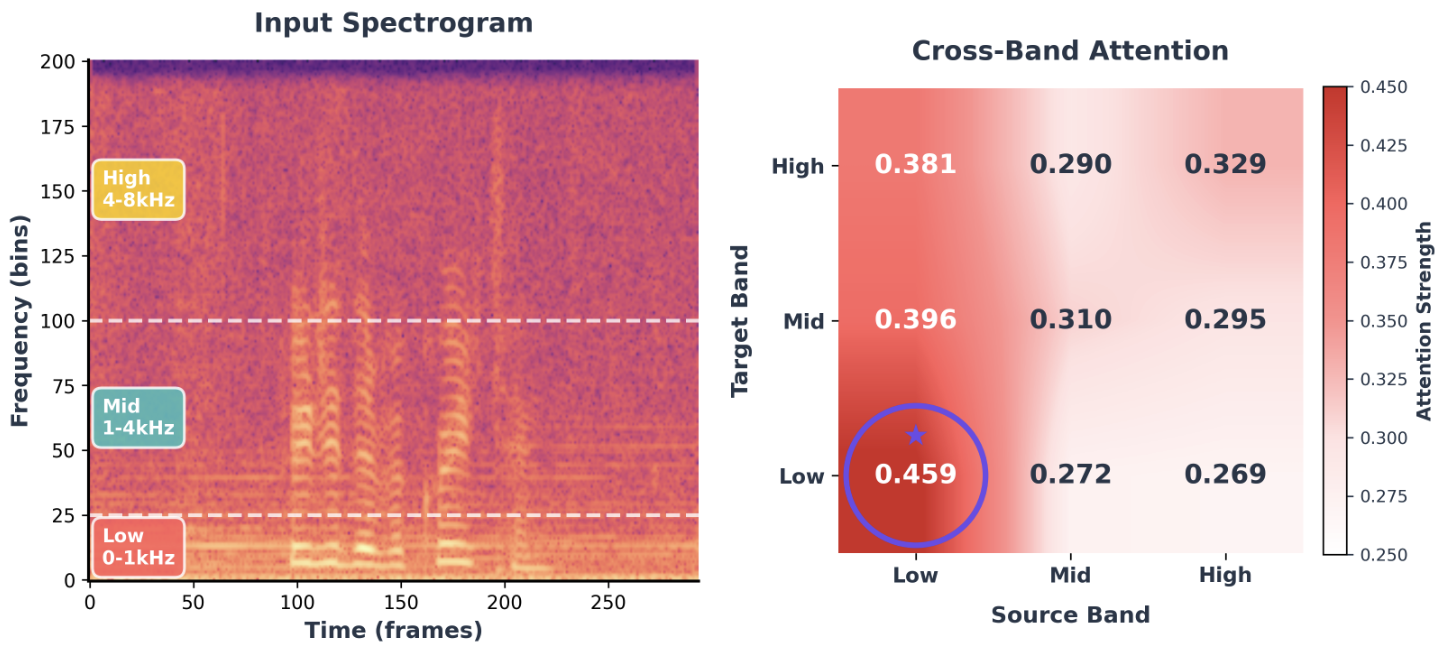}
    \caption{\textbf{Left:} input spectrogram with band boundaries.
    \textbf{Right:} frame-averaged cross-band attention weights
    (Sec.~\ref{sec:crossband}); entry $(i,j)$ denotes the attention
    from band~$i$ to the frequency-pooled summary of band~$j$.}
    \label{fig:attention}
\end{figure}
\vspace{-5mm}
\subsection{Results}
Table~\ref{tab:comparison} compares BASENet-3 against representative methods spanning waveform, complex spectrogram, magnitude--phase, and magnitude--phase--time input representations.\\
\textbf{Quality--efficiency trade-off.}
BASENet-3 achieves PESQ~3.55 with only 0.83\,M parameters and 7.3\,G~MACs---the fewest parameters among all methods reaching PESQ\,$\geq$\,3.50. It matches SE-Mamba~\cite{chao2025investigationincorporatingmambaspeech} (PESQ~3.55) at $4.5\times$ fewer MACs and $2.7\times$ fewer parameters, and surpasses MP-SENet~\cite{lu2023mpsenet} (PESQ~3.50) at $5.1\times$ fewer MACs. Among other lightweight methods, it outperforms DPT-FSNet~\cite{tan2022dptfsnet} ($+0.22$~PESQ, fewer MACs), CMGAN~\cite{cao2022cmgan} ($+0.14$~PESQ at one-third the computation), and MUSE~\cite{lin24h_interspeech} ($+0.18$~PESQ, $+0.15$~CBAK). Compared to BSRNN~\cite{yu23b_interspeech}---which also employs sub-band splitting---BASENet-3 achieves $+0.45$~PESQ with fewer parameters (0.83\,M vs.\ 3\,M), highlighting the benefit of scaled-capacity over uniform sub-band processing.\\
\textbf{Comparison with recent SOTA.}
Mamba-SEUNet~\cite{wang2025mambaseunetmambaunetmonaural} (PESQ~3.57) and MH-SENet~\cite{kim25q_interspeech} (PESQ~3.62) achieve higher PESQ, but at greater cost or with richer inputs: Mamba-SEUNet requires $4.6\times$ more parameters (3.78\,M), while MH-SENet additionally leverages a time-domain input stream (magnitude--phase--time) not available to magnitude--phase methods. Within the same input category, BASENet-3 ties with SE-Mamba for the highest PESQ among magnitude--phase methods while using $4.5\times$ fewer MACs, demonstrating that perceptually scaled capacity is a highly effective inductive bias.\\
\textbf{Causal streaming.}
BASENet-3 natively supports causal inference by replacing the bidirectional GRU with a unidirectional variant---no architectural redesign is required. The causal model (0.81\,M, 7.1\,G~MACs) achieves PESQ~3.44 and STOI~96\%, outperforming several \emph{non-causal} baselines including CMGAN (3.41) and DPT-FSNet (3.33), demonstrating suitability for real-time streaming with only a modest quality reduction ($-0.11$~PESQ).
\subsection{Cross-Band Attention Analysis}
Fig.~\ref{fig:attention} visualises frame-averaged attention weights. The attention matrix $\mathbf{W} \in \mathbb{R}^{B\times B}$ reveals that all bands attend most strongly to the low-frequency band ($W_{\cdot\to\text{low}} \geq 0.381$), confirming that the network exploits fundamental frequency cues for coherent cross-band reconstruction. The low band exhibits the strongest self-attention ($0.459$), reflecting the concentration of harmonic energy, while the mid band shows the most balanced distribution ($0.396/0.310/0.295$), acting as a spectral bridge consistent with the formant region's mediating role in speech perception.
\begin{table}[h]
\centering
\caption{Ablation study on BASENet.}
\label{tab:ablation}
\resizebox{\columnwidth}{!}{%
\begin{tabular}{lccccc}
\toprule
\textbf{Configuration} & \textbf{PESQ} & \textbf{CSIG} & \textbf{CBAK} & \textbf{COVL} \\
\midrule
BASENet-3 & \textbf{3.55} & \textbf{4.65} & \textbf{3.95} & \textbf{4.18} \\
\quad \textit{w/o} freq-adapted & 3.48 & 4.55 & 3.88 & 4.03 \\
\quad \textit{w/o} cross-band attn & 3.42 & 4.43 & 3.81 & 3.89 \\
\quad \textit{w/o} scaled-capacity & 3.44 & 4.45 & 3.85 & 3.97 \\
\midrule
\textbf{BASENet-3} & \textbf{3.55} & \textbf{4.65} & \textbf{3.95} & \textbf{4.18} \\
BASENet-8 & 3.52 & 4.57 & 3.89 & 4.08 \\
BASENet-12 & 3.47 & 4.53 & 3.80 & 3.99 \\
\midrule
\textbf{BASENet-3-CRN} & \textbf{3.55} & \textbf{4.65} & \textbf{3.95} & \textbf{4.18} \\
BASENet-3-MambaTM & 3.53 & 4.62 & 3.94 & 4.17 \\
\bottomrule
\end{tabular}
}
\end{table}
\vspace{-1mm}
\subsection{Ablation Study}
\label{sec:ablation}
Table~\ref{tab:ablation} isolates each component's contribution.\\
\textbf{Architecture components.} Disabling cross-band attention causes the largest degradation ($-0.13$~PESQ), confirming that inter-band information exchange is critical for modelling harmonic relationships. Replacing scaled-capacity allocation with uniform depth yields a comparable drop ($-0.11$~PESQ), validating the density-derived depth rule over equal-depth processing. Removing frequency-adapted processing reduces PESQ by~$0.07$---a smaller gap, suggesting the other components partially compensate, though the full combination remains essential for best performance.\\
\textbf{Band granularity.}
As discussed in Sec.~\ref{sec:overall}, increasing from $B=3$ to $B=8$ and $B=12$ progressively degrades performance ($-0.03$ and $-0.08$~PESQ), suggesting that overly fine partitions limit each branch's spectral context and reduce modelling effectiveness.\\
\textbf{Temporal modelling.}
Replacing the CRN with a Mamba-based temporal module~\cite{chao2025investigationincorporatingmambaspeech} yields comparable performance (PESQ~3.53 vs.\ 3.55), confirming that the architecture's gains stem primarily from the frequency-adapted encoder and cross-band attention rather than the temporal modelling choice.







\section{Conclusion}
We introduced BASENet, a frequency-adaptive speech enhancement network that allocates encoder depth according to Bark-scale critical-band density and restores cross-band coherence via efficient attention, achieving a strong quality--efficiency trade-off on VoiceBank+DEMAND (PESQ~3.55, 0.83\,M parameters) with a causal variant (PESQ~3.44) that surpasses several non-causal baselines. Future work will explore lower-triangular attention masks to exploit the bottom-up harmonic structure of speech.
\section{Generative AI Use Disclosure}
Generative AI tools were used solely for spelling correction 
and sentence reformulation during the preparation of this 
manuscript. No generative AI tool was used to produce 
a significant part of the scientific content, methodology, 
experiments, or results reported in this work.


\bibliographystyle{IEEEtran}
\bibliography{mybib}

@article{moore2012introduction,
  title={An introduction to the psychology of hearing},
  author={Moore, Brian CJ},
  journal={Brill},
  year={2012}
}

@book{loizou2013speech,
  title={Speech enhancement: theory and practice},
  author={Loizou, Philipos C},
  year={2013},
  publisher={CRC press}
}

@inproceedings{yin2020phasen,
  title={{PHASEN}: A phase-and-harmonics-aware speech enhancement network},
  author={Yin, Dacheng and Luo, Chong and Xiong, Zhiwei and Zeng, Wenjun},
  booktitle={Proc. AAAI},
  pages={9458--9465},
  year={2020}
}

@inproceedings{lu2023mpsenet,
  title={{MP-SENet}: A speech enhancement model with parallel denoising of magnitude and phase spectra},
  author={Lu, Ye-Xin and Ai, Yang and Ling, Zhen-Hua},
  booktitle={Proc. Interspeech},
  pages={3834--3838},
  year={2023}
}

@inproceedings{hu2020dccrn,
  title={{DCCRN}: Deep complex convolution recurrent network for phase-aware speech enhancement},
  author={Hu, Yanxin and Liu, Yun and Lv, Shubo and Xing, Mengtao and Zhang, Shimin and Fu, Yihui and Wu, Jian and Zhang, Bihong and Xie, Lei},
  booktitle={Proc. Interspeech},
  pages={2472--2476},
  year={2020}
}

@inproceedings{tan2022dptfsnet,
  title={{DPT-FSNet}: Dual-path transformer based full-band and sub-band fusion network for speech enhancement},
  author={Tan, Ke and Chen, Jitong and Wang, DeLiang},
  booktitle={Proc. ICASSP},
  pages={6857--6861},
  year={2022}
}

@inproceedings{howard2019searching,
  title={Searching for {MobileNetV3}},
  author={Howard, Andrew and Sandler, Mark and Chu, Grace and Chen, Liang-Chieh and Chen, Bo and Tan, Mingxing and Wang, Weijun and Zhu, Yukun and Pang, Ruoming and Vasudevan, Vijay and others},
  booktitle={Proc. ICCV},
  pages={1314--1324},
  year={2019}
}

@inproceedings{hu2018squeeze,
  title={Squeeze-and-excitation networks},
  author={Hu, Jie and Shen, Li and Sun, Gang},
  booktitle={Proc. CVPR},
  pages={7132--7141},
  year={2018}
}

@inproceedings{huang2017densely,
  title={Densely connected convolutional networks},
  author={Huang, Gao and Liu, Zhuang and Van Der Maaten, Laurens and Weinberger, Kilian Q},
  booktitle={Proc. CVPR},
  pages={4700--4708},
  year={2017}
}

@book{rabiner1978digital,
  title={Digital processing of speech signals},
  author={Rabiner, Lawrence R and Schafer, Ronald W},
  year={1978},
  publisher={Prentice-Hall}
}

@inproceedings{tan2019learning,
  title={Learning complex spectral mapping with gated convolutional recurrent networks for monaural speech enhancement},
  author={Tan, Ke and Wang, DeLiang},
  journal={IEEE/ACM Trans. Audio, Speech, Language Process.},
  volume={28},
  pages={380--390},
  year={2019}
}

@inproceedings{valentini2016investigating,
  title={Investigating {RNN}-based speech enhancement methods for noise-robust text-to-speech},
  author={Valentini-Botinhao, Cassia and Wang, Xin and Takaki, Shinji and Yamagishi, Junichi},
  booktitle={Proc. SSW},
  pages={146--152},
  year={2016}
}

@inproceedings{hu2007evaluation,
  title={Evaluation of objective quality measures for speech enhancement},
  author={Hu, Yi and Loizou, Philipos C},
  journal={IEEE Trans. Audio, Speech, Language Process.},
  volume={16},
  number={1},
  pages={229--238},
  year={2007}
}

@inproceedings{defossez2020real,
  title={Real time speech enhancement in the waveform domain},
  author={Defossez, Alexandre and Synnaeve, Gabriel and Adi, Yossi},
  booktitle={Proc. Interspeech},
  pages={3291--3295},
  year={2020}
}

@inproceedings{kim2021seconformer,
  title={{SE-Conformer}: Time-domain speech enhancement using conformer},
  author={Kim, Woo-Jin and Chung, Ji Won and Choi, Yong-Joon and Park, Kyuwoong and Choi, Jung-Won},
  booktitle={Proc. Interspeech},
  pages={2736--2740},
  year={2021}
}

@inproceedings{cao2022cmgan,
  title={{CMGAN}: Conformer-based metric {GAN} for speech enhancement},
  author={Cao, Ruizhi and Abdulatif, Sherif and Yang, Bin},
  booktitle={Proc. Interspeech},
  pages={936--940},
  year={2022}
}

@article{wang2014training,
  title={On training targets for supervised speech separation},
  author={Wang, Yuxuan and Narayanan, Arun and Wang, DeLiang},
  journal={IEEE/ACM Trans. Audio, Speech, Language Process.},
  volume={22},
  number={12},
  pages={1849--1858},
  year={2014}
}

@article{wang2023tfgridnet,
  title={{TF-GridNet}: Integrating full- and sub-band modeling for speech separation},
  author={Wang, Zhong-Qiu and Cornell, Samuele and Choi, Shukjae and Lee, Younglo and Kim, Byeong-Yeol and Watanabe, Shinji},
  journal={IEEE/ACM Trans. Audio, Speech, Language Process.},
  volume={31},
  pages={3221--3236},
  year={2023}
}

@inproceedings{schroter2022deepfilternet,
  title={{DeepFilterNet}: A low complexity speech enhancement framework for full-band audio based on deep filtering},
  author={Schr{\"o}ter, Hendrik and Escalante-B., Alberto N. and Rosenkranz, Tobias and Maier, Andreas},
  booktitle={Proc. ICASSP},
  pages={7407--7411},
  year={2022}
}

@inproceedings{hao2021fullsubnet,
  title={{FullSubNet}: A full-band and sub-band fusion model for real-time single-channel speech enhancement},
  author={Hao, Xiang and Su, Xiangdong and Horaud, Radu and Li, Xiaofei},
  booktitle={Proc. ICASSP},
  pages={6633--6637},
  year={2021}
}

@inproceedings{chen2023intersubnet,
  title={{Inter-SubNet}: Speech enhancement with subband interaction},
  author={Chen, Jun and Rao, Wei and Wang, Zilin and Lin, Jiuxin and Wu, Zhiyong and Wang, Yuxuan and Shang, Shidong and Meng, Helen},
  booktitle={Proc. ICASSP},
  pages={1--5},
  year={2023}
}

@article{richter2023sgmse,
  title={Speech enhancement and dereverberation with diffusion-based generative models},
  author={Richter, Julius and Welker, Simon and Lemercier, Jean-Marie and Lay, Bunlong and Gerkmann, Timo},
  journal={IEEE/ACM Trans. Audio, Speech, Language Process.},
  volume={31},
  pages={2351--2364},
  year={2023}
}

@inproceedings{liu2024speechflow,
  title={Generative pre-training for speech with flow matching},
  author={Liu, Alexander H. and Le, Matt and Vyas, Apoorv and Shi, Bowen and Tjandra, Andros and Hsu, Wei-Ning},
  booktitle={Proc. ICLR},
  year={2024}
}

@inproceedings{rix2001pesq,
  title={Perceptual evaluation of speech quality ({PESQ})---a new method for speech quality assessment of telephone networks and codecs},
  author={Rix, Antony W and Beerends, John G and Hollier, Michael P and Hekstra, Andries P},
  booktitle={Proc. ICASSP},
  volume={2},
  pages={749--752},
  year={2001}
}

@article{taal2011stoi,
  title={An algorithm for intelligibility prediction of time--frequency weighted noisy speech},
  author={Taal, Cees H and Hendriks, Richard C and Heusdens, Richard and Jensen, Jesper},
  journal={IEEE Trans. Audio, Speech, Language Process.},
  volume={19},
  number={7},
  pages={2125--2136},
  year={2011}
}

@inproceedings{lin24h_interspeech,
  title     = {{MUSE: Flexible Voiceprint Receptive Fields and Multi-Path Fusion Enhanced Taylor Transformer for U-Net-based Speech Enhancement}},
  author    = {Zizhen Lin and Xiaoting Chen and Junyu Wang},
  year      = {2024},
  booktitle = {{Interspeech 2024}},
  pages     = {672--676},
  doi       = {10.21437/Interspeech.2024-1017},
  issn      = {2958-1796},
}

@inproceedings{Shin_2022, series={interspeech\_2022},
   title={Multi-View Attention Transfer for Efficient Speech Enhancement},
   url={http://dx.doi.org/10.21437/Interspeech.2022-10251},
   DOI={10.21437/interspeech.2022-10251},
   booktitle={Interspeech 2022},
   publisher={ISCA},
   author={Shin, Wooseok and Park, Hyun Joon and Kim, Jin Sob and Lee, Byung Hoon and Han, Sung Won},
   year={2022},
   month=sep, pages={1198–1202},
   collection={interspeech_2022}
}

@inproceedings{kingma2017adam,
title={Adam: A Method for Stochastic Optimization},
author={Diederik P. Kingma and Jimmy Ba},
booktitle={International Conference on Learning Representations},
year={2015},
url={http://arxiv.org/abs/1412.6980},
}

@inproceedings{yu23b_interspeech,
  title     = {{High Fidelity Speech Enhancement with Band-split RNN}},
  author    = {Jianwei Yu and Hangting Chen and Yi Luo and Rongzhi Gu and Chao Weng},
  year      = {2023},
  booktitle = {{Interspeech 2023}},
  pages     = {2483--2487},
  doi       = {10.21437/Interspeech.2023-1433},
  issn      = {2958-1796},
}

@article{zwicker1961subdivision,
    author = {Zwicker, E.},
    title = {Subdivision of the Audible Frequency Range into Critical Bands (Frequenzgruppen)},
    journal = {The Journal of the Acoustical Society of America},
    volume = {33},
    number = {2},
    pages = {248-248},
    year = {1961},
    month = {02},
    issn = {0001-4966},
    doi = {10.1121/1.1908630},
    url = {https://doi.org/10.1121/1.1908630},
    eprint = {https://pubs.aip.org/asa/jasa/article-pdf/33/2/248/18743398/248_1_online.pdf},
}

@article{traunmuller1990analytical,
  title={Analytical expressions for the tonotopic sensory scale},
  author={Traunm{\"u}ller, Hartmut},
  journal={The journal of the acoustical society of America},
  volume={88},
  number={1},
  pages={97--100},
  year={1990},
  publisher={Acoustical Society of America}
}

@inproceedings{kim25q_interspeech,
  title     = {{Mamba-based Hybrid Model for Speech Enhancement}},
  author    = {Se-Ha Kim and Tae-Gyeong Kim and Chang-Jae Chun},
  year      = {2025},
  booktitle = {{Interspeech 2025}},
  pages     = {5163--5167},
  doi       = {10.21437/Interspeech.2025-1476},
  issn      = {2958-1796},
}

@inproceedings{wang2025mambaseunetmambaunetmonaural,
  title={Mamba-SEUNet: Mamba UNet for monaural speech enhancement},
  author={Wang, Junyu and Lin, Zizhen and Wang, Tianrui and Ge, Meng and Wang, Longbiao and Dang, Jianwu},
  booktitle={ICASSP 2025-2025 IEEE International Conference on Acoustics, Speech and Signal Processing (ICASSP)},
  pages={1--5},
  year={2025},
  organization={IEEE}
}

@inproceedings{chao2025investigationincorporatingmambaspeech,
  title={An investigation of incorporating mamba for speech enhancement},
  author={Chao, Rong and Cheng, Wen-Huang and La Quatra, Moreno and Siniscalchi, Sabato Marco and Yang, Chao-Han Huck and Fu, Szu-Wei and Tsao, Yu},
  booktitle={2024 IEEE Spoken Language Technology Workshop (SLT)},
  pages={302--308},
  year={2024},
  organization={IEEE}
}

\end{document}